\documentclass[prl,twocolumn,superscriptaddress,longbibliography]{revtex4-2}
\usepackage{amsmath,amssymb,bm}
\usepackage{graphicx}
\usepackage{epstopdf}
\usepackage{latexsym}
\usepackage{subfigure}
\usepackage[usenames, dvipsnames]{color}
\usepackage[usenames, dvipsnames]{xcolor}
\usepackage[linkbordercolor=red]{hyperref}
\usepackage{natbib}
\usepackage{braket}
\usepackage{float}
\usepackage[normalem]{ulem}
\usepackage{comment}
\usepackage{mathtools}
\usepackage{array}
\usepackage{tabu}
\usepackage{multirow}
\usepackage{chemformula}
\usepackage{svg}
\usepackage{booktabs} 

\newcommand\redsout{\bgroup\markoverwith{\textcolor{red}{\rule[0.5ex]{2pt}{0.4pt}}}\ULon}
\newcommand\bluesout{\bgroup\markoverwith{\textcolor{blue}{\rule[0.5ex]{2pt}{0.4pt}}}\ULon}

\newcommand{\prlsec}[1]{{{\textit{#1}:}}} 

\usepackage{tikz}
\newcommand{\myProj}{%
\protect\tikz[baseline]{\protect\draw[line width=0.4mm] (0,0.205) arc (180:360:0.1) (0.2,-0.045) arc (0:180:0.1)}%
}

\newcommand{\SPhide}[1]{{}}
\newcommand{\SKuhide}[1]{{}}
\newcommand{\JDhide}[1]{{}}

\begin{document}
\title{Pseudocriticality in antiferromagnetic spin chains}
\author{Sankalp Kumar}
\affiliation{Department of Physics, Indian Institute of
Technology Bombay, Mumbai, MH 400076, India}
\author{Sumiran Pujari}
\email{sumiran.pujari@iitb.ac.in}
\affiliation{Department of Physics, Indian Institute of
Technology Bombay, Mumbai, MH 400076, India}
\affiliation{Max Planck Institute for the Physics of Complex Systems, 01187 Dresden, Germany}
\author{Jonathan D'Emidio}
\email{jdemidio@utk.edu}
\affiliation{Department of Physics and Astronomy, University of Tennessee, Knoxville, TN 37996, USA}

\begin{abstract}
Weak first-order pseudocriticality with approximate scale invariance has been observed in a variety of settings, including the intriguing case of deconfined criticality in 2+1 dimensions.
Recently, this has been interpreted as extremely slow flows (``walking behavior") for real-valued couplings in proximity to a {\em bona fide} critical point with complex-valued couplings, described by a complex conformal field theory (CFT).
Here we study an SU($N$) generalization of the the Heisenberg antiferromagnet, which is a familiar model for deconfined criticality in 2+1 dimensions.  We show that in 1+1 dimensions the model is located near a complex CFT, whose proximity can be tuned as a function of $N$.
We employ state-of-the-art quantum Monte Carlo simulations for \emph{continuous} $N$ along with an improved loop estimator for the R\'{e}nyi entanglement entropy based on a nonequilibrium work protocol.
These techniques allow us to track the central charge of this model in detail as a function of $N$, where we observe excellent agreement with CFT predictions.  Notably, this includes the region $N>2$, where the CFT moves into the complex plane and pseudocritical drifts enable us to recover the real part of the complex central charge with remarkable accuracy.
Since the present model with $N=3$ is also equivalent to the spin-1 biquadratic model, our work sheds new light on the dimerized phase of the spin-1 chain, demonstrating that it is pseudocritical and proximate to a complex CFT.
\end{abstract}
\maketitle
\newpage

The understanding of critical phenomena, enabled by the development of the renormalization group (RG)~\cite{Wilson_1974,Wilson_1975,Wilson_1983} in conjunction with conformal field theory~\cite{Polyakov_1970,Zomolodchikov_1986,Polchinski_1988}, is an astounding achievement in the fields of statistical and condensed matter physics. 
The notion of universality is central to this understanding. 
Vastly different systems display the same universal scaling laws when brought to the verge of a second-order phase transition, or critical point, where the correlation length diverges. 
In contrast, the behavior at first-order transitions, with a finite correlation length, depends on microscopic details of the system and has therefore attracted less interest.

Recently, however, interest has gathered around weak first-order transitions where a large, but finite, correlation length may be explained by proximity to a critical point located outside of the physical parameter space~\cite{Chong_Wang_etal_PRX_2017,Gorbenko_etal_JHEP_2018,Gorbenko_etal_Scipost_2018}. 
The 2D classical Potts model~\cite{Wu_RevModPhys_1982}, a generalization of the Ising model with $Q$-component spins, is the primary example of this phenomenon. 
When $Q\leq 4$ the model contains a second-order transition, where the magnetization vanishes continuously as a function of temperature, whereas the transition shows a weak first-order discontinuity for $Q\gtrsim 4$. 
This is understood more generally as arising from the merger and annihilation of the critical and tri-critical Potts RG fixed points occurring at $Q=4$, which produces a pair of fixed points with complex values of the temperature and (vacancy) fugacity for $Q>4$~\cite{Gorbenko_etal_Scipost_2018}. 
Importantly, for real parameters near $Q\gtrsim 4$, the presence of nearby complex fixed points results in extremely slow RG flows at intermediate length scales, manifesting as a weakly first-order transition with approximate scale invariance.

Numerical work has confirmed this picture for extensions of the 2D classical~\cite{Jacobsen_Wiese_PRL_2024} and 1+1D quantum Potts model~\cite{Ma_He_PRB_2019,Tang_et_al_2024}, as well as the closely related 2D classical O($n$) loop model~\cite{Haldar_etal_PRL_2023}. 
In particular, density matrix renormalization group (DMRG) calculations of the quantum Potts chain above $Q>4$, in the weakly first-order regime, revealed a scale-dependent value of the central charge that drifts toward zero at large sizes~\cite{Ma_He_PRB_2019}.  
Importantly, the form of the drift is determined from the RG flow equations and allows for an accurate determination of the {\em real part} of the central charge of the nearby complex CFT.

We note that for the well-known models of 2+1D deconfined quantum critical points~\cite{Senthil_et_al_2004,Sandvik_JQ}, which separate distinct symmetry-broken ordered phases, drifting critical exponents have been commonly observed~\cite{Kawashima_et_al_2013,Nahum_etal_PRX_2015,Demidio_2023}. 
This has led to the proposal of pseudocriticality, arising from proximity to a complex CFT, as a possible mechanism for drifting in models of deconfined criticality~\cite{Chong_Wang_etal_PRX_2017,Ma_Wang_PRB_2020,Zhou_et_al_2024}. 
Another possibility is that the deconfined critical point exists as a multicritical point with real couplings but suffers from a sign problem~\cite{Takahashi_et_al_2024}. 
This situation motivates a better understanding of pseudocritical drifts in basic models of quantum magnetism, especially beginning in one dimension where rigorous connections can be made with exactly-known results from CFT.

In this work, by studying finite-size drifts of the central charge, we establish complex-CFT-induced pseudocriticality in a simple 1+1D quantum spin model that extends beyond the Potts case. 
This model is intimately connected with the physics of quantum antiferromagnets, contains the Heisenberg model as a special case, and is a natural starting point for studying deconfined criticality in 2+1D.  
It also contains the spin-1 biquadratic chain as a special case in the pseudocritical regime. 
This reveals that the dimerized phase of the spin-1 chain is in close proximity to a complex CFT.

Our results are based on state-of-the-art quantum Monte Carlo (QMC) techniques using a loop representation of our SU($N$) spin chain partition function. 
This provides two crucial advantages that make our study possible.  
Firstly, it allows us to extend the symmetry group of the SU($N$) chain to real values of $N>0$~\cite{Beach_2009,Resummation}. 
Thus, we \emph{continuously} tune the distance from the complex-CFT for $N>2$, revealing a large window of approximate conformal scaling.  
Secondly, the loop representation allows us to develop a highly efficient algorithm for computing the R\'enyi entanglement entropy (EE), which combines recently developed techniques based on nonequilibrium work~\cite{Alba_PRE_2017,Jon_Noneqlbm} along with a new loop-based estimator.  
Precise calculations of the R\'enyi EE give us direct access to finite-size estimates of the central charge, which we study in detail throughout the critical and pseudocritical regimes. 
Interestingly, we also find that subsystem oscillations of the R\'enyi EE decay with a power given by the Potts thermal exponent.

\begin{figure}[t]
    \includegraphics[width=0.95\linewidth]{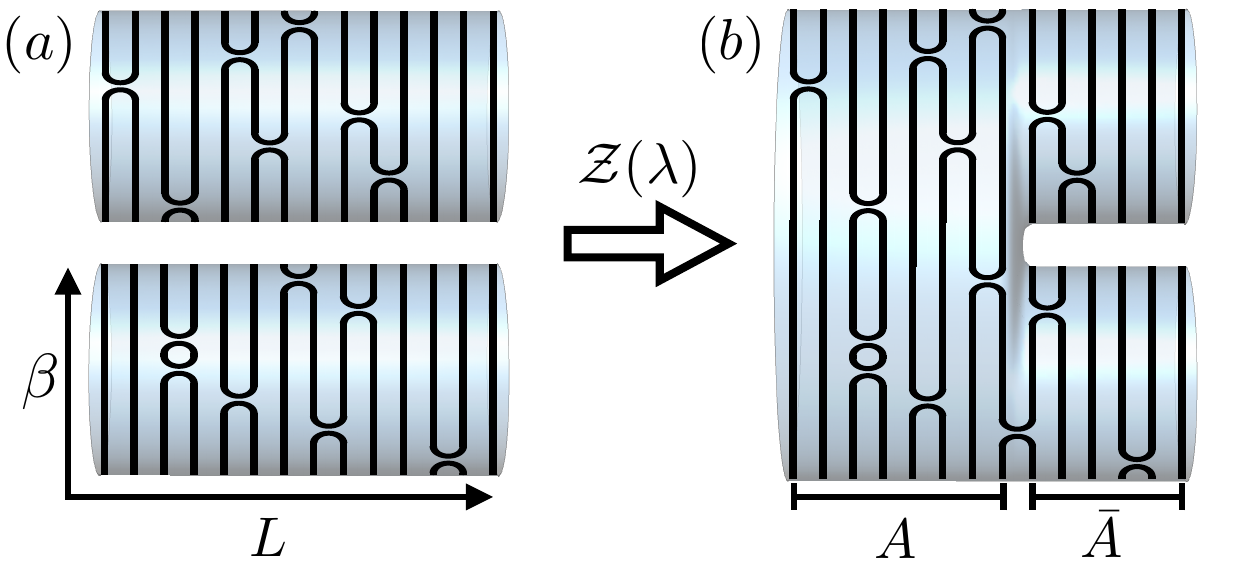}
    \caption{\label{fig:loops} (a) A configuration $c$ of the ensemble $Z_{\varnothing}$, consisting of a product of bond operators $\myProj \equiv \sum_{\alpha,\beta} |\alpha_i \alpha_{i+1}\rangle\langle\beta_i\beta_{i+1}|$ that form closed loops due to the trace structure. The QMC weight of this configuration is given by  $w_c N^{\ell^{\varnothing}_c}$, where $\ell^{\varnothing}_c$ is the number of loops. (b) The same configuration in the ensemble $Z_{A}$ with a weight given by $w_c N^{\ell^{A}_c}$. In this work we introduce the interpolating ensemble $\mathcal{Z}(\lambda)=\sum_c w_c N^{\lambda \ell^{A}_c + (1-\lambda)\ell^{\varnothing}_c}$ such that $\mathcal{Z}(0)=Z_{\varnothing}$ and $\mathcal{Z}(1)=Z_{A}$. This ensemble allows for highly precise calculations of the R\'enyi EE using a nonequilibrium work protocol.}
\end{figure}

\prlsec{Model} 
We consider a well-known SU($N$) generalization of the Heisenberg antiferromagnetic spin chain~\cite{Affleck_1985,Read_PRl_1989,Read_NucPhy_1989,Read_1990,Read_1991} using the fundamental representation of SU($N$) on one sublattice and the conjugate to the fundamental on the other sublattice. In this representation a singlet can be formed on two adjacent sites, written as $\tfrac{1}{\sqrt{N}}\textstyle\sum^N_{\alpha=1}|\alpha_i\alpha_{i+1}\rangle$. The Hamiltonian is simply given as minus the sum over all nearest-neighbor singlet projectors:
 \begin{equation}
H = -\frac{J}{N}\sum^L_{i=1}\sum^N_{\alpha,\beta=1} |\alpha_i\alpha_{i+1}\rangle\langle \beta_i\beta_{i+1}|,
\label{eq:hamProj}
\end{equation}
where $\alpha,\beta,\ldots$ stands for the color or flavor index of the SU($N$) degrees of freedom.

When $N=2$ this model is equivalent, after a sublattice rotation, to the spin-$\tfrac{1}{2}$ Heisenberg antiferromagnet and features a power-law correlated, gapless groundstate described by the $O(3)$ nonlinear sigma model with topological $\theta=\pi$ term~\cite{Haldane_1981,Haldane_1983a,Haldane_1983b} with central charge $c=1$~\cite{Affleck_NucPhysB_1986,Affleck_Haldane_PRB_1987}.  For $N=3$ this model is equivalent to the spin-1 biquadratic model $H_{bq}=-J\textstyle\sum_{i} (\mathbf{S}_i \cdot \mathbf{S}_{i+1})^2$ which features a gapped groundstate residing in the dimerized phase of the familiar bilinear-biquadratic model~\cite{Barber_Batchelor_1989,Klumper_EPL_1989,xian_1993,Lauchli_PRB_2006}.  For larger $N$, this model has a rich history in describing quantum antiferromagnets via perturbative $1/N$ expansions ~\cite{Affleck_1985,Read_PRl_1989,Read_NucPhy_1989,Read_1990,Read_1991}. In two dimensions, this model ~\cite{Beach_2009,Song_etal_PRB_2024} and similar variants ~\cite{Ribhu_SUN,Block_PRL_2013,Jon_easyplane,Kaul_Melko_Sandvik_review_2013} naturally house deconfined quantum phase transitions between antiferromagnetic and valence bond solid ground states when $N$ reaches a critical value.  For example, $N_c\approx 4.6$ on the two-dimensional square lattice ~\cite{Ribhu_SUN,Beach_2009,Song_etal_PRB_2024}.

The present SU($N$) model in 1+1D can be mapped to the quantum Potts chain with $Q=N^2$, as shown by Affleck~\cite{Affleck_1990}. 
Furthermore, the critical points realized in these models are equivalent to those of the classical O($n$) loop models~\cite{Nienhuis_1982,Nienhuis_1991,Haldar_etal_PRL_2023,Blote_fullypacked_PRL_1994}. 
In fact, it was pointed out in Ref.~\cite{Jacobsen_Wiese_PRL_2024} that a similar quantum spin chain, expressed in terms of Temperley-Lieb generators, should realize a complex-CFT. 
We also point out that the SU($N$) chain automatically lives on a critical-pseudocritical line (parametrized by $N$)~\cite{footnote_dimerization_field}, whereas the quantum Potts chain needs to be tuned to the transition point.

\prlsec{QMC Technique}
We now describe our algorithm for computing the R\'enyi EE of the ground state of Eq. (\ref{eq:hamProj}) for continuous values of $N$, which gives us direct access to the central charge.  We refer the reader to Refs.~\cite{Beach_2009,Resummation} for standard QMC simulations of this model with continuous $N$. This formulation allows for highly precise loop estimators of physical quantities. These estimators take advantage of spin symmetry to greatly reduce the stochastic error. We note that a loop estimator of the R\'enyi EE was first developed for ground state projector simulations~\cite{Hasting_etal_PRL_2010} and recently extended to finite-temperature simulations~\cite{Song_etal_PRB_2024}. Here, we develop a much more efficient~\cite{suppinfo} nonequilibrium loop protocol that builds on recent advancements in computing R\'enyi EE in standard QMC simulations~\cite{Alba_PRE_2017,Jon_Noneqlbm,Zhao_etal_review_2022}. This idea is similar to a recent approach developed in ground state projector simulations~\cite{Zhou_etal_PRB_2024}.

The second R\'enyi EE is defined as $S_2 = -\ln(\text{Tr}(\rho^2_A))$, where $\rho_A$ is the reduced density matrix of a region $A$: $\rho_A = \mathrm{Tr}_{\bar{A}}(e^{-\beta H})/\mathrm{Tr}(e^{-\beta H})$. Here we use $\beta$ sufficiently large so that only the ground state contributes~\cite{suppinfo}. $S_2$ can also be expressed as a free energy difference between two generalized partition functions as $S_2 = -\ln(\frac{Z_A}{Z_{\varnothing}})$~\cite{Calabrese2004,Calabrese_2009}, where $Z_{X}\equiv\text{Tr}_X(\text{Tr}_{\bar{X}}(e^{-\beta H})^2)$, and $\varnothing$ refers to the empty set. The generalized partition functions have a loop decomposition in the resummed stochastic series expansion that is expressed as a sum over configurations: $Z_{X}=\sum_c w_c N^{\ell^X_c}$, where the weight factor $w_c$ depends on the expansion order~\cite{Sandvik_comp_stud} and is independent of $X$ and the spin degeneracy factor $N^{\ell^X_c}$ depends on the number of loops formed by the trace condition over $X$, denoted as $\ell^X_c$.  Importantly, $Z_A$ and $Z_{\varnothing}$ differ only in their degeneracy factors, since the number of loops depends on the trace condition, with $\ell^{\varnothing}_c \geq \ell^{A}_c$, see Fig. (\ref{fig:loops}).

We now formulate a highly efficient loop-based protocol for computing $S_2$ by introducing an interpolating ensemble 
\begin{equation}
\mathcal{Z}(\lambda)=\sum_c w_c N^{\lambda \ell^A_c + (1-\lambda)\ell^{\varnothing}_c}
\label{eq:Zlam}
\end{equation}
such that $\mathcal{Z}(0)=Z_{\varnothing}$ and $\mathcal{Z}(1)=Z_{A}$.  Following~\cite{Alba_PRE_2017,Jon_Noneqlbm}, $S_2$ can be computed via a nonequilibrium simulation of $\mathcal{Z}(\lambda)$ where $\lambda$ is varied from 0 to 1.  In each nonequilibrium process the corresponding work is calculated as
\begin{equation}
W = \ln(N)\int^{t_f}_{t_i} \dot{\lambda}(\ell^{\varnothing}_{c(t)}-\ell^{A}_{c(t)}) dt
\label{eq:Work}
\end{equation}
where $\lambda(t_i)=0$ and $\lambda(t_f)=1$.  Jarzynski's equality~\cite{Jarzynski_1997} then provides us with a statistically exact formula to compute $S_2$, with $e^{-S_2} = \langle e^{-W}\rangle$. The average is computed over independent nonequilibrium trajectories 

As a final note, we have found that the strongly dimerized region of our model for $N>2$ poses challenges for efficient QMC sampling.  This relates to long tunneling times required to move between the two dimerization patterns of the chain. To circumvent this issue we have developed a replica shifting method that substantially improves our data quality in this regime~\cite{suppinfo}.

\begin{figure}[t]
    \includegraphics[width=1.0\linewidth]{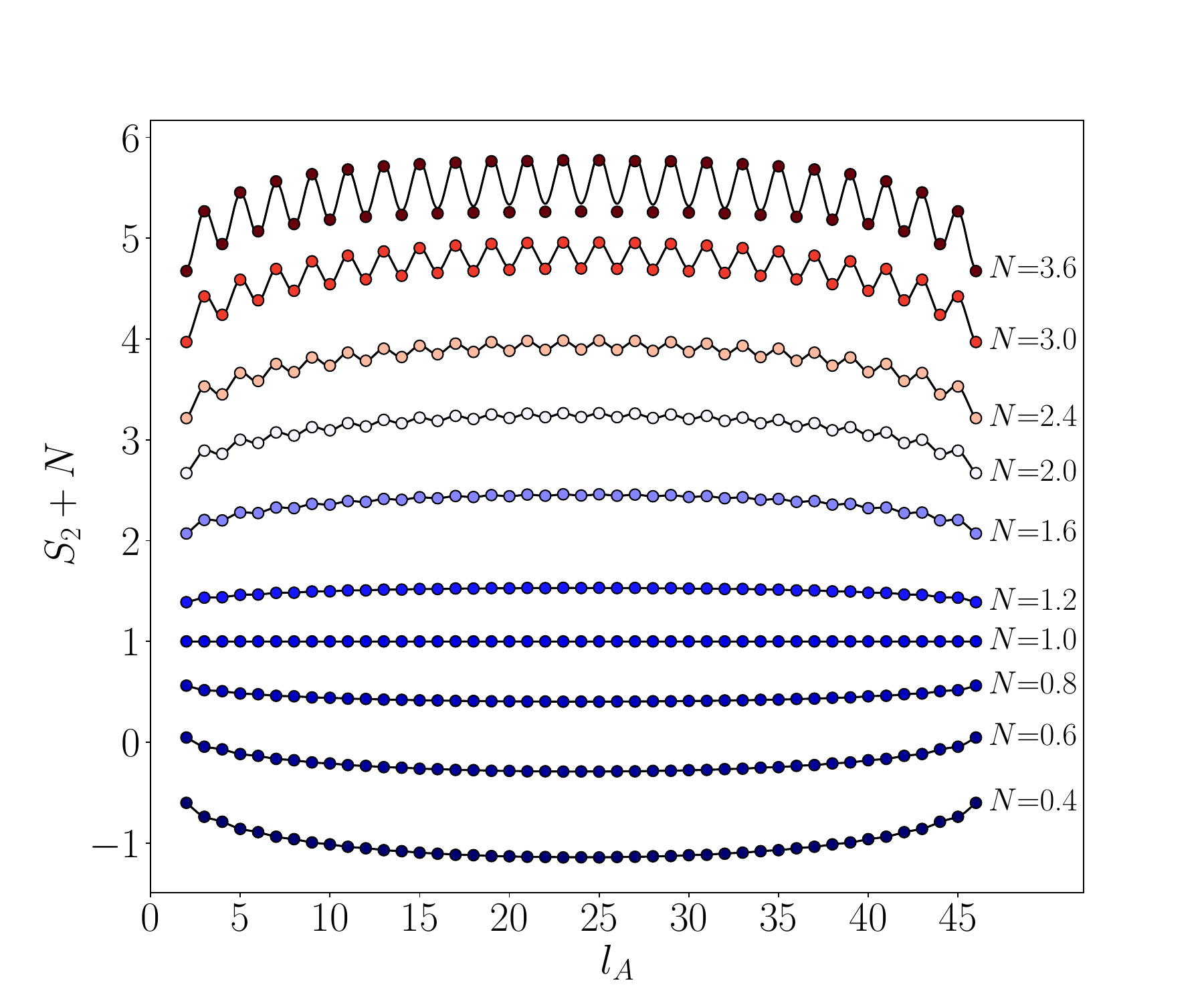}
    \caption{\label{fig:calabrese_cardy_fit} R\'enyi EE as a function of subsystem size ($l_{A}$) for a chain of length $L=48$ with periodic boundary conditions for different values of $N$. Colored points are obtained from QMC (where we have ommited $l_A=1,L-1$ and shifted the data by $N$ for clarity). Solid black lines are the fit to scaling form in Eq. (\ref{eq:CC_reltn}).  The numerically fitted central charges and and exponent are shown in Fig. (\ref{fig:main_plot}c).}
\end{figure}

\begin{figure*}
    \centering
    \begin{minipage}[t]{0.5\textwidth}
        \centering
        \includegraphics[height=7cm]{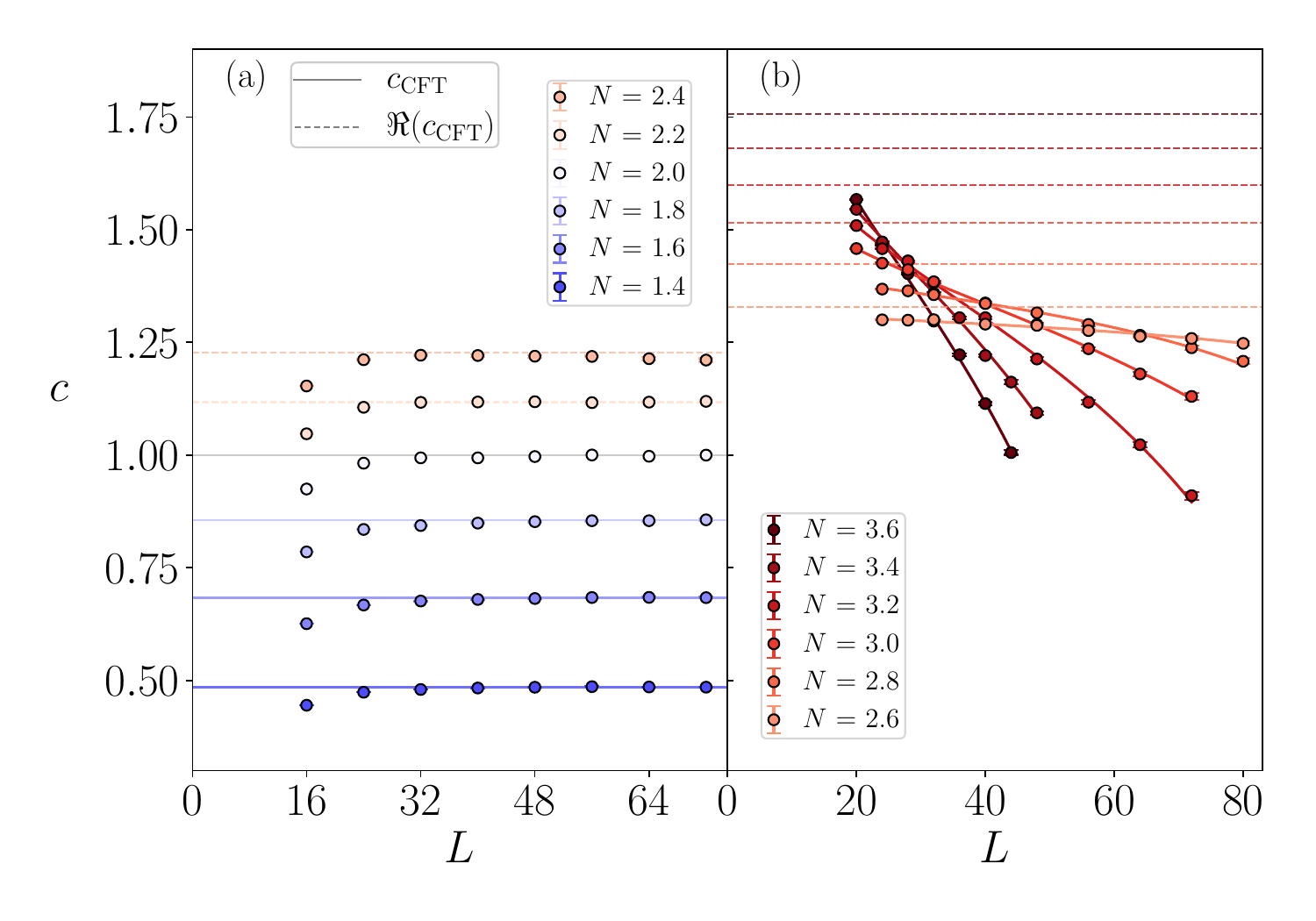}
    \end{minipage}%
    \hfill
    \begin{minipage}[t]{0.5\textwidth}
        \centering
        \includegraphics[height=7cm]{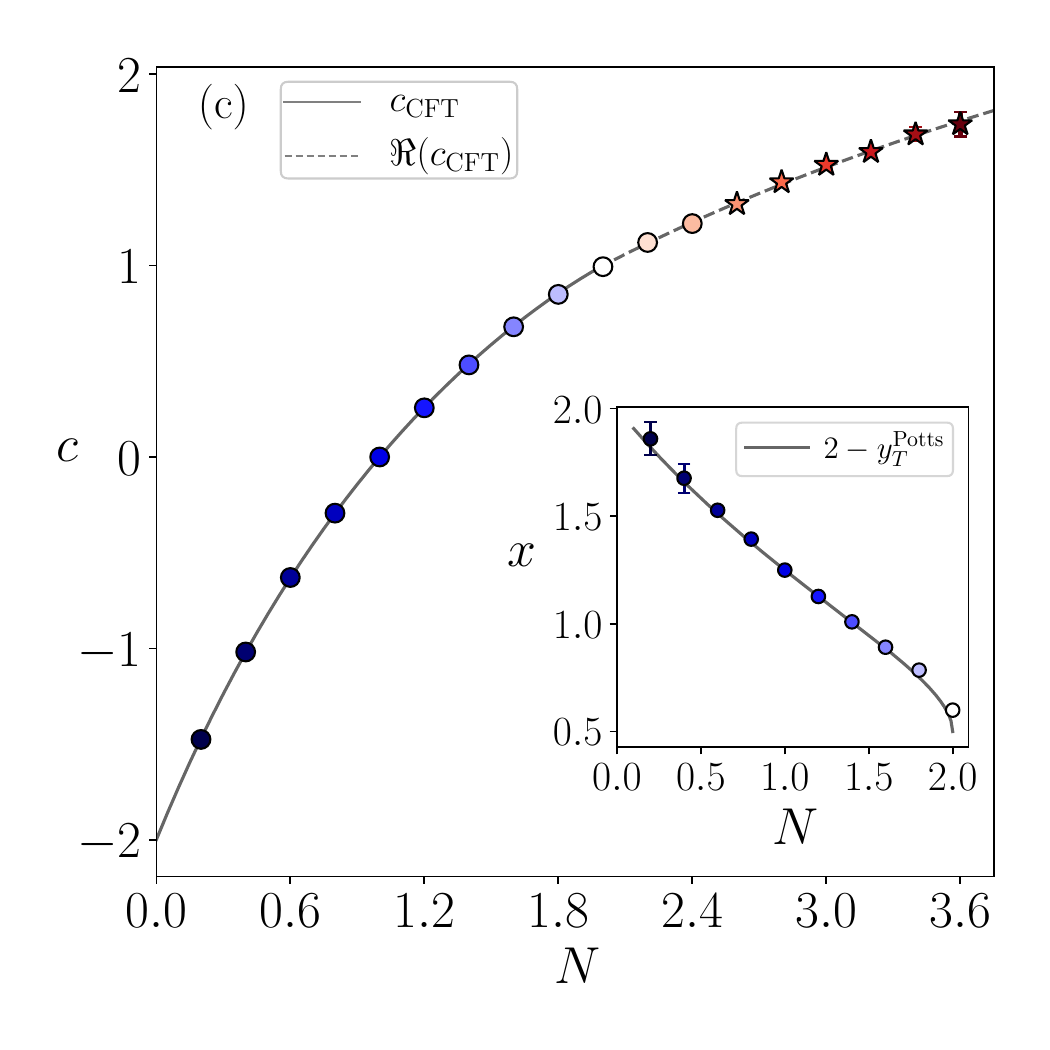}
    \end{minipage}
    \caption{(a,b) The central charge extracted from chains of total size $L$ for various values of $N$.  Solid horizontal lines are the central charge given by Eq. (\ref{eq:CFT_prediction}). When $N>2$ the central charge becomes complex, in which case we show the real part as dashed horizontal lines. In (a) we show the regime where direct convergence to the CFT value is observed, which includes $N=2.2,2.4$, even though $c$ ultimately flows to zero in our model when $N>2$. In (b) we show the drift dominated regime, where $c$ differs substantially from CFT predictions, making a direct comparison impossible. However, the form of the drift in Eq. (\ref{eq:c_drift}) can be used to fit (solid curves on the plot) to the CFT value. (c) The extracted values of the central charge compared with the CFT prediction (curved line). The stars are obtained by the drift formula fitting, whereas circles are based on finite-size converged values in (a) (we also include additional data at smaller $N$).  The inset shows the decay exponent associated with subsystem oscillations of the R\'enyi EE, which closely match two minus the Potts thermal exponent in Eq. (\ref{eq:scaling_dim}).
    }
    \label{fig:main_plot}
\end{figure*}

\prlsec{Results}
In Fig.(\ref{fig:calabrese_cardy_fit}) we show our $S_2$ data as a function of subsystem size for periodic chains of length $L=48$ and $N\in [0.4,3.6]$.  For gapless antiferromagnetic spin chains, we expect a scaling form given by~\cite{Holzhey1994,Vidal2003,Calabrese2004,calabrese_etal_2010}
\begin{equation}
S_2(l_A) = \frac{c}{4} \ln D(l_A,L) + f\cos(\pi l_A) D(l_A,L)^{-x} +b
\label{eq:CC_reltn}
\end{equation}
where $D(l_A,L)=\frac{L}{\pi}\sin(\frac{\pi l_A}{L})$ is the chord distance. The logarithmic term is characteristic of gapless spin chains described by a CFT with central charge $c$. Additionally, there is a subleading oscillatory piece that arises in $S_2$ due to strong antiferromagnetic correlations. Importantly, the oscillations contain additional information about the CFT, namely the decay exponent $x$, which, for instance, in the XXZ model is given by the Luttinger parameter~\cite{calabrese_etal_2010,Berganza_JStatMech_2012}.

There are several important qualitative features of our data in Fig. \ref{fig:calabrese_cardy_fit}. Firstly, $S_2(l_A)$ changes from concave up to concave down at $N=1$ and is exactly zero at $N=1$ since $Z_A$=$Z_{\varnothing}$ there. This indicates that the central charge changes sign at this point, and $c=0$ when $N=1$. Secondly, as $N$ increases, we observe stronger oscillations due to a greater tendency to form dimers at larger $N$. For all $N>2$ the chain is dimerized in the thermodynamic limit~\cite{Affleck_1990}, but our finite-size data still closely follows the scaling form in Eq. (\ref{eq:CC_reltn}), as evidenced by the numerical fits represented by the solid lines in Fig. (\ref{fig:calabrese_cardy_fit}). Only at our largest value of $N=3.6$ in Fig. (\ref{fig:calabrese_cardy_fit}) do we see obvious deviations away from the scaling form near $l_A \approx L/2$. In what follows, in order to avoid any spurious effects while fitting to Eq. (\ref{eq:CC_reltn}) in the pseudocritical regime, we instead opt for a more robust extraction of $c$ when $N>2$ based on a subtraction scheme that cancels out the effect of oscillations~\cite{suppinfo}.

Moving to a quantitative analysis of our data, Fig. (\ref{fig:main_plot}a) shows the convergence of $c(L)$, the central charge extracted from chains of size $L$.  The solid horizontal lines are the predicted values from CFT given by the formula~\cite{,diFrancesco_etal_JstatP_1987}
\begin{equation}
c = 1-\frac{6(1-g)^{2}}{g} 
   \quad\text{and}\quad 
N=-2\cos(\pi g),
\label{eq:CFT_prediction}
\end{equation}
expressed in terms of the Coulomb gas coupling constant $g\in[\tfrac{1}{2},1]$.
This is the same central charge as in the Potts and $O(n)$ loop models.  
Importantly, this formula gives a complex-valued central charge when $N>2$. 
However, for $2<N\lesssim2.4$ our finite-size estimates $c(L)$ converge almost perfectly to the real part $\Re(c)$ (dashed horizontal lines in Fig. (\ref{fig:main_plot}a), while they are expected to eventually drift to zero at much larger sizes. 
This eventual drift is just barely noticeable for the case of $N=2.4$ in Fig.~\ref{fig:main_plot}(a) at the largest sizes.
The agreement with the CFT is again depicted in Fig. (\ref{fig:main_plot}c), where we have also plotted the agreement found in the region $N<1$ with a negative central charge. 
The inset of Fig. (\ref{fig:main_plot}c) includes the value of the decay exponent $x$, extracted from fits to Eq. (\ref{eq:CC_reltn}) for $N\leq 2$. 
We find remarkable agreement with $x=2-y^{\text{Potts}}_T$, where $y^{\text{Potts}}_T$ is the Potts thermal exponent given by~\cite{Nijs_1979}
\begin{equation}
y^{\text{Potts}}_T =\tfrac{3}{2}[2+\pi/(\mu-\pi)] 
   \quad\text{and}\quad 
N=2\cos(\mu).
    \label{eq:scaling_dim}
\end{equation}
This is an interesting case of unusual corrections to scaling of the R\'enyi EE~\cite{calabrese_etal_2010,Cardy_unusual_2010}, with an exponent different than the Luttinger parameter.  The same exponent can also be expressed as $x=2\Delta_{1,2}$, with $\Delta_{p,q}$ given by the Kac formula~\cite{Belavin1984,Friedan1984,Blote_1989}. We note that slight deviations of $x$ from the CFT value are observed near $N=2$ due to the presence of known logarithmic corrections~\cite{Affleck_1989}. For $N\gtrsim2$, even before reaching the drifting regime, we can no longer match $x$ to a CFT form, indicating a possible breakdown of the oscillatory form in Eq. (\ref{eq:CC_reltn}) for complex scaling dimensions.

Finally we move to the drift-dominated regime shown in Fig. (\ref{fig:main_plot}b) for values of $N\geq2.6$. This regime is characterized by a monotonically decreasing $c(L)$ from small to larger size chains. We see that these values deviate significantly from the real part of Eq. (\ref{eq:CFT_prediction}) (dashed lines), making a direct comparison impossible. Recently, however, it was shown in the Potts model with $Q>4$ that the real part of the CFT central charge can be recovered based on finite-size drifts~\cite{Ma_He_PRB_2019}. Using the RG flow of the central charge near the Potts complex CFT, one finds the size-dependent form
\begin{equation}
    c(L) = c_{R} - \alpha \tan (\alpha^{1/3} \log(L/L_{0})),
\label{eq:c_drift}
\end{equation}
where $c_R,\alpha,L_0$ are taken as fit parameters. Here $c_R$ is the real part of the complex CFT central charge.  We note that Eq. (\ref{eq:c_drift}) is derived from the lowest-order RG flow equations and, in principle, is only valid very close to the complex-CFT. We in fact observe remarkable agreement with Eq. (\ref{eq:c_drift}) over a surprisingly wide range of $N>2$. In Fig. (\ref{fig:main_plot}b) the solid curved lines are fits of our numerical data to Eq. (\ref{eq:c_drift}).  The resulting fitted values of $c_R$ are plotted as stars in Fig. (\ref{fig:main_plot}c).  We see that even for our largest value of $N=3.6$, which corresponds to a $Q\approx 13$ Potts model, we can accurately recover the real part of the complex CFT central charge.

\prlsec{Conclusions}
We demonstrated complex CFT driven pseudocritical behavior in a simple and well-known SU($N$) extension of the antiferromagnetic Heisenberg spin chain. 
By studying entanglement with statistically exact QMC simulations for continuously varying values of $N$, we demonstrated excellent agreement with CFT predictions.
In addition to the region $N>2$ where we see the effect of complex-valued central charges, we also observed a region of negative real central charges when $N<1$.
Interestingly, and further connecting entanglement with critical phenomena, we found that the power-law decay of the subsystem oscillations of the R\'enyi EE is governed by the Potts thermal exponent.

Our work opens several avenues for further numerical and theoretical investigations. 
Firstly, we believe that an in-depth study of the dimerized spin-1 bilinear-biquadratic chain is now motivated, given the proximate complex CFT found at $N=3$. 
In particular, it would be interesting to extract finite-size estimates of the CFT spectrum and study drifts of the scaling dimensions.  
It may also be desirable to consider non-Hermitian extensions of the spin-1 model that actually contain the complex CFT in the phase diagram. 
Finally, it is worth emphasizing that the present model is described by the $\mathbb{CP}^{N-1}$ field theory~\cite{DAdda_NuclPhysB_1978,Read_PRl_1989} in the continuum. 
Our work thus provides a direct impetus to investigate complex CFTs in 1+1D (and higher-dimensional) field theories with topological terms.

\section{Acknowledgements}
We thank Anders W. Sandvik for helpful comments.
SP acknowledges support from
SERB-DST, India via Grant No. MTR/2022/000386 and 
partially by Grant No. CRG/2021/003024. 
SP is grateful to the strongly correlated 
electron systems group at Laboratory of Theoretical
Physics, University Paul Sabatier, Toulouse, France
for the hospitality during parts of this
work. JD acknowledges support from the National Science
Foundation (NSF) under award No. OSI-232680.

\bibliography{refs}

\begin{thebibliography}{73}
\bibitem [{foo()}]{footnote_dimerization_field}%
  \BibitemOpen
  \href@noop {} {}\bibinfo {note} {Formally, the transition is with respect to the following dimerization or valence bond symmetry breaking field $h \sum_{i}\sum^N_{\alpha,\beta=1} (-1)^i |\alpha_i\alpha_{i+1}\rangle\langle \beta_i\beta_{i+1}|$}\BibitemShut {NoStop}%
\end{thebibliography}

\section{Supplemental material}

\subsection{Benchmarking with Exact Diagonalization}

\begin{figure}[h]
    \includegraphics[width=1.0\linewidth, height = 8.0cm]{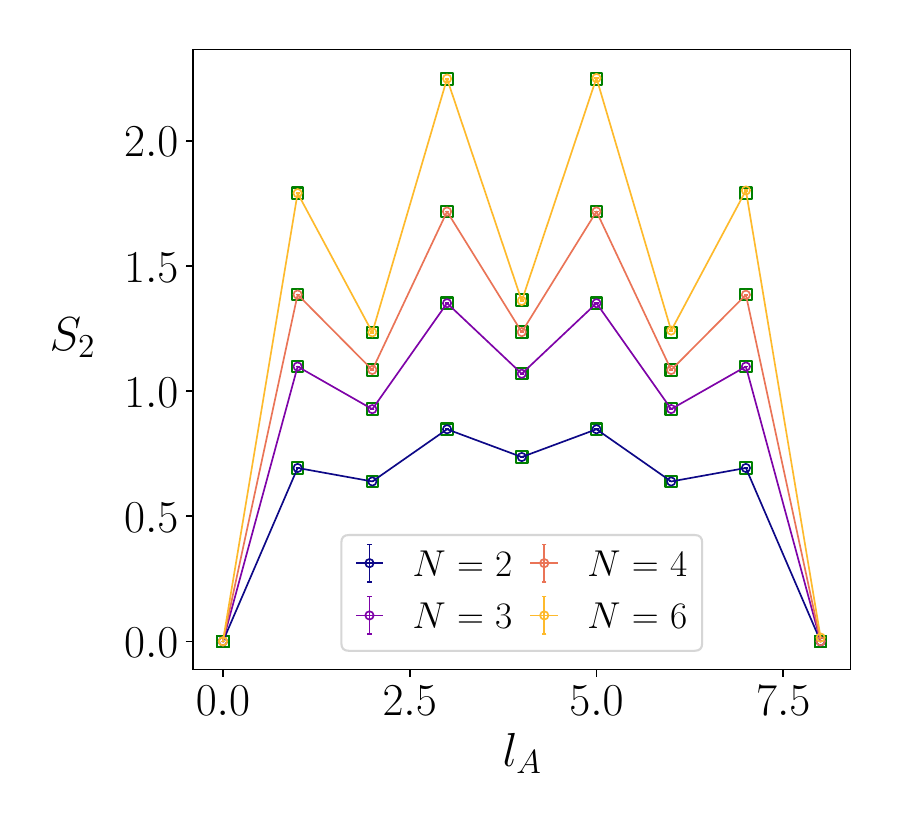}
    \caption{\label{fig:ED_benchmarking}
    We benchmark the computed REE ($S_{2}$) using our RSSE-based QMC method (colored circle data points connected by lines) with the exact diagonalization values (green square data points) for a periodic chain of length $L=8$ and for $N=2,3,4,6$. We find perfect agreement between QMC and exact diagonalization for all values of $N$.
    }
\end{figure}

To validate our QMC implementation, we benchmark the 2nd R\'enyi entanglement entropy (REE) against exact diagonalization (ED). In Fig.~\ref{fig:ED_benchmarking}, we compare the REE obtained from QMC and ED for a periodic chain of length $L=8$ and for $N=2,3,4,6$. We find excellent agreement between the two methods across all values of $N$, confirming the accuracy of our QMC approach.

\begin{figure}[h]
    \includegraphics[width=1.0\linewidth]{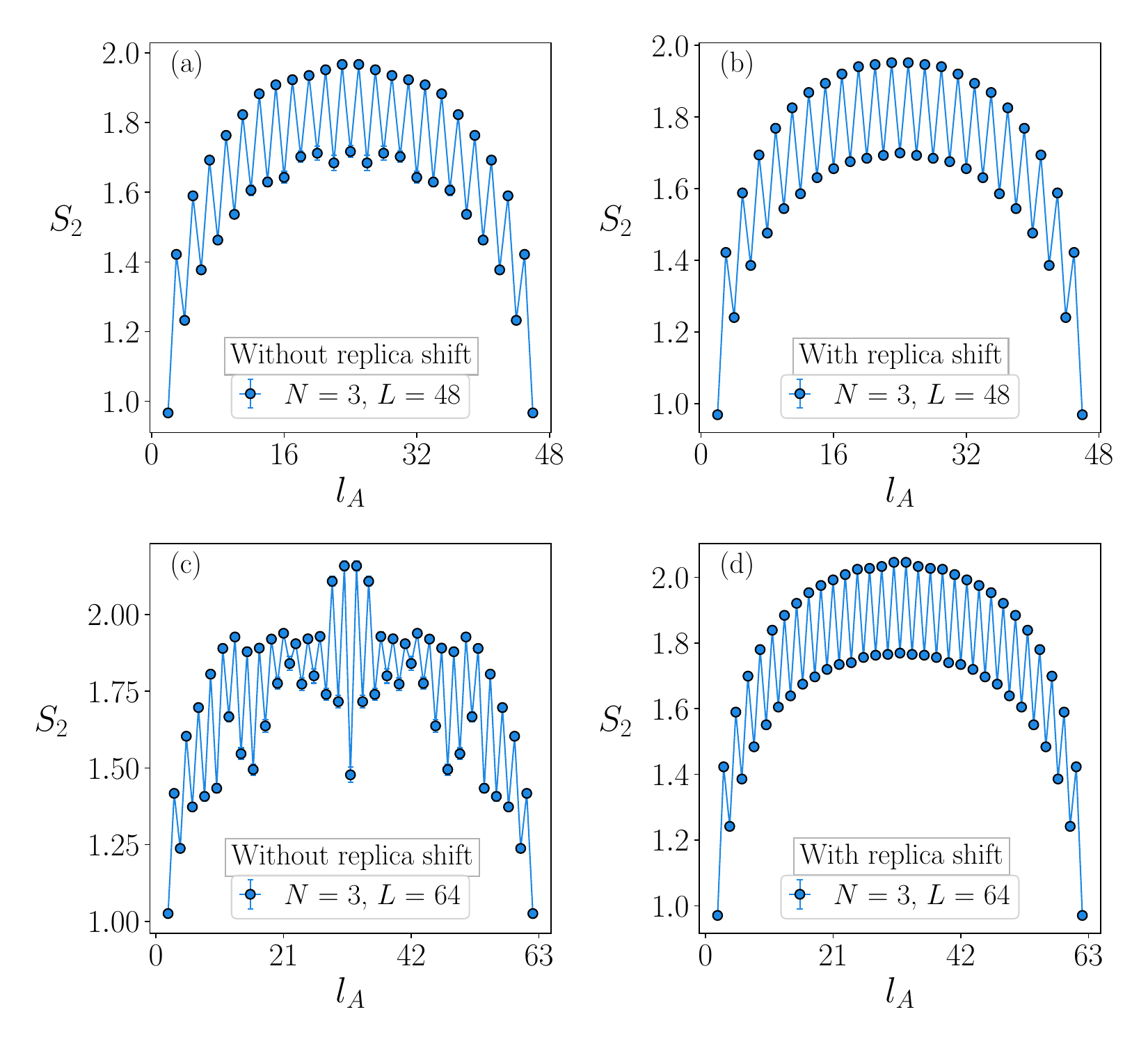}
    \caption{\label{fig:replica_shift} Comparison between the REE ($S_{2}$) as a function of the subsystem size ($l_{A}$) computed using with and without replica shift method for $N=3$. In (a,c), we have the REE without the replica shift for $L=48,64$. As $L$ increases, the tunneling time required to move between the two dimerization patterns of the chain also increases. As a result, we can not compute REE accurately. On the other hand, with the additional update of the replica shift, we can compute REE accurately, even for higher system sizes, as shown in (b,d).}
\end{figure}

\begin{figure*}[t]
    \includegraphics[width=1.0\linewidth]{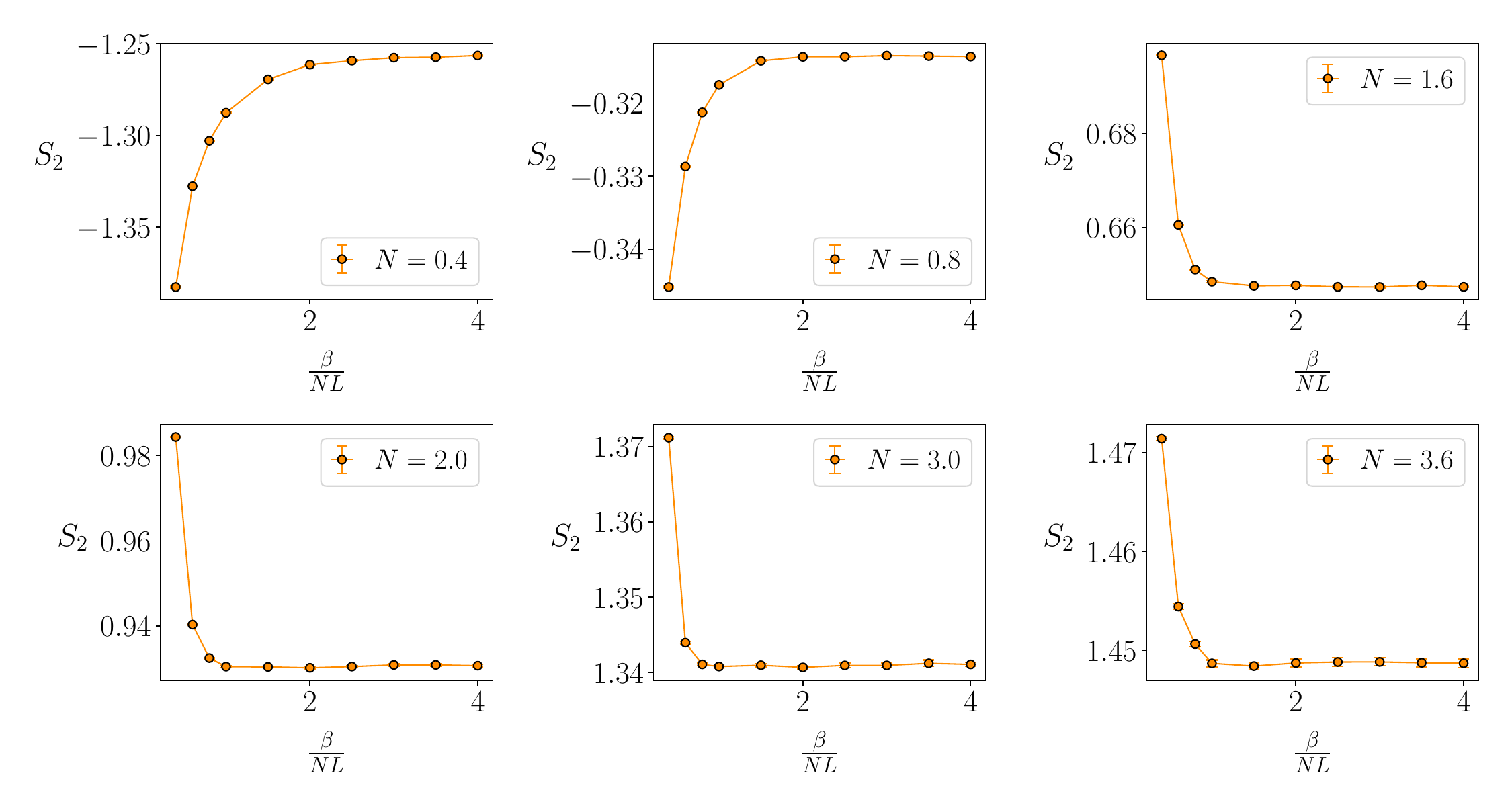}
    \caption{\label{fig:beta_convergence}
    In order to probe the ground state in a finite temperature QMC methods, it is necessary to choose sufficiently large $\beta$ value, where $\beta = \frac{1}{T}$ is the inverse temperature. Here we show the convergence of the half system REE of a periodic $L=16$ site chain as function of $\frac{\beta}{NL}$ for various $N$ values. For smaller $N$, it converges at much higher values of $\frac{\beta}{NL}$ compared to larger $N$. Based on this convergence, we have chosen $\frac{\beta}{NL}= x$, where $x=4$ for $0<N<0.5 $, $x=3$ for $0.5<N<1.0$, $x=2$ for $1.0<N<2.0$, and $x=1$ for $N\geq 2.0$.
    }
\end{figure*}

\subsection{ Replica shift update for $N>2$}

As discussed in the main text, for $N>2$, the ground state is in the  dimerized phase. For a periodic chain, two kinds of dimer coverings are possible and standard RSSE loop updates suffer ergodicity problems while transitioning from one type of dimer covering to another. To solve this problem, during the nonequilibrium work measurement, additional to the standard RSSE loop updates, we use an extra update termed as ``replica shift update" where we  shift all the bond-operators in the operator-string by one lattice unit, either to the left or to the  right (with equal probability), and we accept this move with a probability depending on the weight ratio. When we shift the operator-string by one lattice unit, total number of loops in a configurations may change, as a result the weight of the configuration also changes. 

We will now outline the algorithm for the replica shift.  In what follows, we assume that we are computing the free energy difference $-\log(Z_{A_2}/Z_{A_1})$, where $A_2$ and $A_1$ are arbitrary sets of lattice sites.
we first equilibrate the system in the ${Z_{A_1}}$ ensemble of configurations (i.e. at $\lambda = 0$). To compute the nonequilibrium work done in transitioning from the ${Z_{A_1}}$ ensemble to the $Z_{A_2}$ ensemble the parameter $\lambda$ is incrementally increased from $0$ to $1$. After each increment of  $\lambda$, an additional ``replica shift update" is performed prior to measuring the corresponding work increment.

(1) For each replica, decide with probability $0.5$ whether to attempt a bond shift left or right.

(2) Once the direction is chosen, accept the shift with a probability equal to the ratio of the weight of the shifted configuration to that of the original configuration:

\begin{align}
    P_{\text{shift}} &= \frac{W_{c_{\text{shifted}}}}{W_{c_{\text{original}}}} = N^{(1-\lambda)\Delta \ell_{A_1}} N^{\lambda\Delta \ell_{A_2}}
\end{align}
where $W_c$ is the full QMC weight of a configuration and $\Delta \ell_{X}$ is the change in the total number of loops in a configuration corresponding to the $Z_{X}$ partition function after the replica shift update.

In Fig.~\ref{fig:replica_shift}, we show the computed REE ($S_{2}$)  with and without this additional replica shift update at $N=3$ for a periodic chain of size $L=48$ and $L = 64$. We find that with this additional update we can compute $S_2$ more accurately.

\subsection{Ground state convergence}

Since we employ a finite-temperature QMC method, it is crucial to work at sufficiently low temperatures to access ground state properties. In Fig.~\ref{fig:beta_convergence}, we plot the half-system REE of a periodic 16-site chain as a function of $\beta / (NL)$ for various values of $N$. We observe that for smaller $N$, the REE converges at significantly higher values of $\beta / (NL)$ compared to larger $N$. Based on this trend, we choose the inverse temperature as $\beta = xNL$, where $x=4$ for $0 < N < 0.5$, $x=3$ for $0.5 \leq N < 1.0$, $x=2$ for $1.0 \leq N < 2.0$, and $x=1$ for $N \geq 2.0$.


\subsection{Comparison with the SSE method}

In the resummed stochastic series expansion (RSSE)-QMC the partition functions have a loop decomposition that is expressed as a sum over loop configurations: $Z=\sum_c w_c N^{\ell_c}$, where the weight factor $w_c$ only depends on the expansion order and $N^{\ell_c}$ is the spin degeneracy which depends on the total number of loops in configuration~\cite{Resummation}. In other words, each loop configuration in the RSSE corresponds to $N^{\ell_c}$ spin configurations in the SSE. This means that RSSE can sample spin configurations more efficiently than SSE, as demonstrated by the shorter autocorrelation times observed in RSSE-QMC compared to conventional SSE-QMC~\cite{Resummation}. This motivates the expectation that a nonequilibrium work method based on RSSE will outperform the SSE-based method when computing the REE.

\begin{figure}[h]
    \centering
    \begin{minipage}[b]{0.5\textwidth}
        \centering
        \includegraphics[height=7cm]{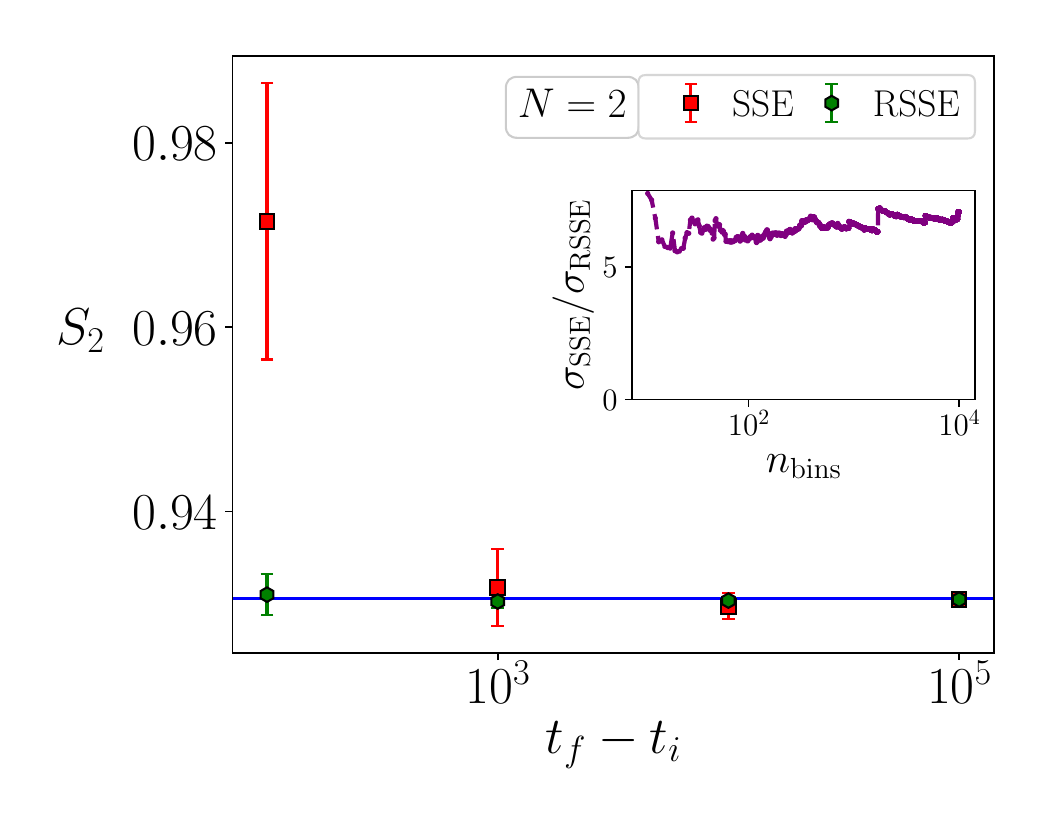}
    \end{minipage}%
    \hfill
    \begin{minipage}[b]{0.5\textwidth}
        \centering
        \includegraphics[height=7cm]{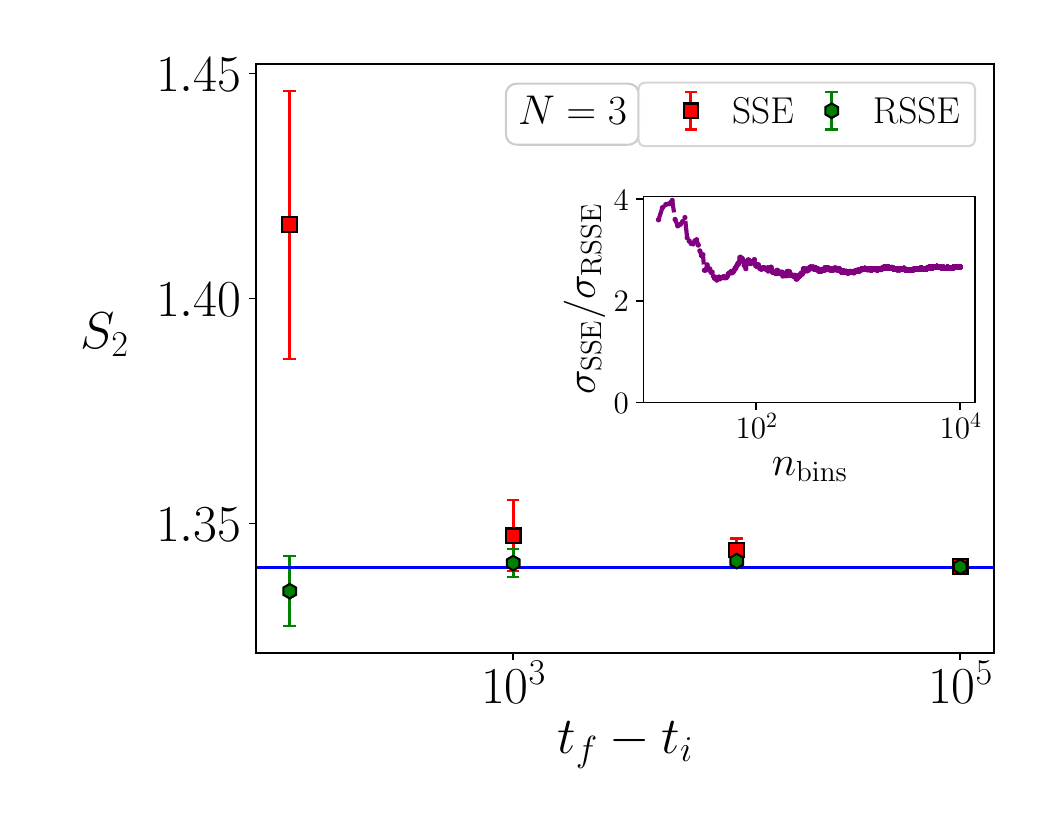}
    \end{minipage}
    \caption{Comparison between the SSE and RSSE method. We compute the half chain entanglement entropy of a periodic chain of length $L=16$ for $N=2$ and $3$ as a function of the quench duration ($t_f-t_i$). At faster quenches (i.e., smaller $t_f-t_i$), the results from the two methods differ. However, as the quench becomes slower, they converge to the same value, indicated by the horizontal blue line in the figure. In the inset, we show the ratio of the error bars for both methods at $t_f-t_i=10^{5}$, where both give the same mean value.
    }
    \label{fig:SSE_RSSE}
\end{figure}

To test this, we compare the nonequilibrium work method for computing the REE using both SSE-QMC and RSSE-QMC. We compute the half-chain entanglement entropy for a periodic chain of length $L = 16$ with $N = 2$ and $3$, as a function of the quench duration $(t_f - t_i)$ as shown in Fig.~\ref{fig:SSE_RSSE}. At short quench durations (i.e., fast quenches), the results obtained from the two methods differ appreciably. However, as the quench becomes slower, both methods converge to the same value, indicated by the horizontal blue line in the figure. We use this convergence point as a reference to assess the relative performance of the two methods across various quench rates.
We find that the RSSE-based method consistently outperforms the SSE-based method over the entire range of quench rates. Notably, RSSE produces significantly smaller error bars and approaches the correct asymptotic value even at relatively fast quench rates. The inset of the figure shows the ratio of error bars for the two methods at $t_f - t_i = 10^{5}$, where both methods give the same mean value but RSSE-based method has much smaller error bars compared to the SSE-based method.

\subsection{Extracting central charge and scaling dimension from the  REE :}

\begin{figure*}[t]
    \includegraphics[width=1.0\linewidth]{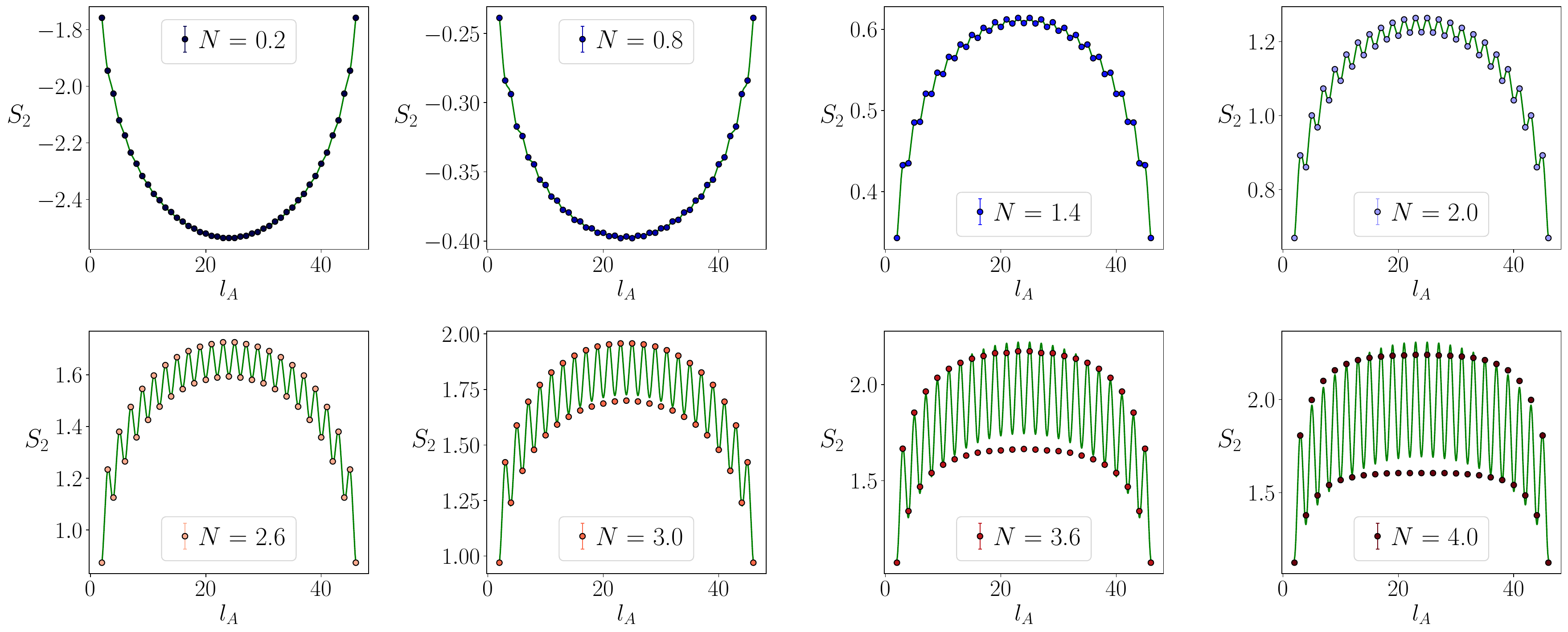}
    \caption{\label{fig:CC_fit}
    Computed REE ($S_{2}$) for chain of size $L=48$ with periodic boundary conditions as a function of subsystem size ($l_{A}$) for $N = 0.2,0.8,1.4,2.0,2.6,3.0,3.6,4.0$. Solid colored data points are the QMC values and green solid lines are the fits to the Calabrese–Cardy scaling form. We have an excellent agreement between the Calabrese–Cardy scaling form and the REE data starting from the $N=0.2$, where REE is negative to up to $N=2.6$. For much higher $N$, where the system is more gapped we start to see the disagreement for larger $l_{A}$ values, as can be seen for $N=3.0,3.6,4.0$. But for the smaller susbsystem sizes Calabrese–Cardy scaling form is still a very good fit and we can extract the central charge using the REE value for smaller $l$ values. Another interesting observation is that for $N>2.6$, where the disagreement starts, the Calabrese–Cardy scaling form fits well for the upper arc compared to lower arc, as clearly visible for $N=3.0$ and also for $N=3.6$ and $4.0$. For $N=4.0$, the system is strongly dimerized which results in the REE getting flat as the $l_{A}$ increases. 
    }
\end{figure*}

The REE of a block subsystem of size $l_A$ in a gapless critical spin chain of size $L$ with periodic boundary conditions follows the Calabrese-Cardy scaling form~\cite{Holzhey1994,Vidal2003,Calabrese2004,calabrese_etal_2010}: 

\begin{equation}
    S_{2}(l_A) = S_{\text{log}}(l_A) + S_{\text{osc}}(l_A) + b
    \label{eq.CC_form}
\end{equation}

where
 \begin{equation}
     S_{\text{log}}(l_A) =  \frac{c}{4} \ln \left[ \frac{L}{\pi} \sin \left( \frac{\pi l_A}{L} \right) \right] 
 \end{equation}

and 
\begin{equation}
    S_{\text{osc}}(l_A) = F \cos(\pi l_A) \left[ \frac{2L}{\pi} \sin \left( \frac{\pi l_A}{L} \right) \right]^{-x}.
\end{equation}
We have included the oscillatory piece of the R\'enyi EE that arises due to strong antiferromagnetic correlations. Here $c$ is the central charge, 
$F$ is a universal scaling function (which is a constant in this case), and $x$ is a universal decay exponent.

For $N\leq 2$ we extract the central charge and the decay exponent from the REE data by doing a fit to the Eqn. (\ref{eq.CC_form}). In table~\ref{Table:N_2.4} we list all the extracted values of the central charge ($c_{\text{QMC}}$) and the decay exponent ($x_{\text{QMC}}$), along with the corresponding CFT prediction ($c_{\text{CFT}}$ and $x_{\text{CFT}}$) for comparison. We find that the extracted values are in excellent agreement with the CFT predictions, within the statistical error.

For $N>2$, since the ground state is no longer critical, the REE will begin to deviate from Eq. (\ref{eq.CC_form}). For small $L$ the REE still follows the gapless scaling form. However, as $L$ increases, disagreement appears around the subsystem size $l_A=\frac{L}{2}$.
Due to this limitation, for $N > 2$, we employ an alternative approach to extract the central charge, referred to here as the subtraction method. This method is based on the following principle: for a $L$ site chain, the REE difference between its two subsystems of size $l_{A_1}$ and $l_{A_2}$ can be computed as:
\begin{align}
    \Delta S^{l_{A_2},l_{A_1}} &= S_{2}(l_{A_2}) - S_{2}(l_{A_1})\\
        &= \Delta S_{\text{log}}^{l_{A_2},l_{A_1}} + \Delta S_{\text{osc}}^{l_{A_2},l_{A_1}}\\
        &= -\ln\bigg(\frac{Z_{A_2}}{Z_{A_1}}\bigg)
\end{align}

If we take these pair of subsystems of length $(l_{A_1},l_{A_2})$ such that both are either even or odd, then $\Delta S_{\text{osc}}^{l_{A_2},l_{A_1}}$ is very small and we can neglect this term. Then we can compute the central charge as:

\begin{equation}
    c^{l_{A_2},l_{A_1}} = \frac{4}{\ln\Big(\frac{\sin (\pi l_{A_2}/L)}{\sin (\pi l_{A_1}/L)}\Big)}\Delta S_{\text{log}}^{l_{A_2},l_{A_1}}
\end{equation}

 Now, if we take two such subsystem pairs, one where both subsystems are of even length and another of odd length.
 The  oscillatory term $\Delta S_{\text{osc}}^{l_{A_2},l_{A_1}}$ for both such pairs is of approximately the same magnitude but different in sign:

\begin{equation}
    \Delta S ^{\mathrm{even,even}}_{\text{osc}} \approx - \Delta S ^{\mathrm{odd,odd}}_{\text{osc}}
\end{equation}
This makes the total contribution from both neglected oscillatory terms approximately zero.
The final central charge is taken as the average of the two values.

 \begin{equation}
     c(L) = \frac{c^{\mathrm{even,even}} + c^{\mathrm{odd,odd}}}{2}
 \end{equation}

In Fig.~\ref{fig:CC_fit}, we plot the REE as a function of subsystem size for a periodic chain with $48$ sites, considering various values of $N \in (0, 4]$. The REE data is fitted to the Calabrese–Cardy scaling form (Eq.~\ref{eq.CC_form}), with the resulting fit curves overlaid on the original data. Deviations from the Calabrese–Cardy scaling form begin to emerge for $N > 2.6$. For instance, at $N = 3$, a noticeable discrepancy appears near $L/2$. Even at larger values such as $N = 3.6$ and $N = 4.0$, the Calabrese–Cardy scaling form continues to describe the REE well for small subsystem sizes but breaks down around $L/2$.

To extract the central charge, we consider two pairs of subsystem lengths: $(L/4, L/4 - 2)$ and $(L/4 - 1, L/4 - 3)$, with $L/4$ chosen to be even. These sizes are centered around $L/4$ because, for $N > 2.0$, the ground state's gapped nature is expected to have minimal influence at smaller subsystem sizes. Since the deviation from the Calabrese–Cardy scaling form is most pronounced near $L/2$, this choice allows for a reliable extraction of the central charge across all values of $N$. Other subsystems around $L/3$, $L/5$, $L/8$, etc., can also be considered.


\subsection{Central charge drift analysis}

As discussed in the main text, in the dimerized phase ($N>2$), the finite size central charge ($c(L)$) has a downward drift as $L$ increases and it will become zero in the thermodynamic limit. Computed $c(L)$ values for $N\in[2.6,3.6]$ are listed in Table~\ref{Table:C_L}.
We can extract the real part of the complex central charge using the following central charge drift relation~\cite{Ma_He_PRB_2019}:


\begin{equation}
    c(L) = c_{R} - \alpha \tan (\gamma \ln (L/L_{0}))
\end{equation}
where $c_{R}, \alpha , \gamma, L_{o}$ are the fit parameters.
$c_{R}$ is the real part of the complex CFT central charge.
For the Potts model $\alpha$ and $\gamma$ are related to each other as $\alpha = \gamma^{3}$. We do a three-parameter fit to extract $c_{R},\alpha, L_{o}$.

\begin{equation}
    c(L) = c_{R} - \alpha \tan (\alpha^{1/3} \ln (L/L_{0}))
    \label{eq:c_drift}
\end{equation}
Also note that for a given set of values of $c_{R},\alpha$, many $L_{0}$ values will fit the drift relation~\ref{eq:c_drift}. These $L_{0}$ values are related to each other as follows:

\begin{equation}
    \label{eq:L0}
  \Tilde{L_{0}} = \exp\bigg(\ln L_{0} \pm \frac{(2n+1)\pi}{\alpha^{1/3}}\bigg) \text{, where } n=0,\pm1,\pm2, ...
\end{equation}

 We use the following method to asses systematic errors in the fitting: for a dataset with $M$ points, we systematically leave out one data point at a time, perform the fit on the remaining $M-1$ points, and extract the fitting parameters. This procedure is repeated for each data point.
 Finally, we compute the average values of the fit parameters and estimate their errors using the standard error formula, based on the fluctuations across the different fits. Final values of the fit parameters are listed in Table~\ref{Table:N_3.6}.

 We note that $L_0$, as interpreted as the RG scale that coincides with the fixed point, should decrease with $N$.  We, however, observe the opposite effect, even though $L_0$ cannot be uniquely defined, as in Eq. (\ref{eq:L0}). We note that other fit parameters in the drift formula have previously shown somewhat counterintuitive behavior~\cite{Ma_He_PRB_2019}.  Despite this, we find robust values for the fitted value of $c_R$ in all cases.


\subsection{ Non-unitary Violation of the $c-d$ conjecture}

\begin{figure}[t]
    \includegraphics[width=1.0\linewidth]{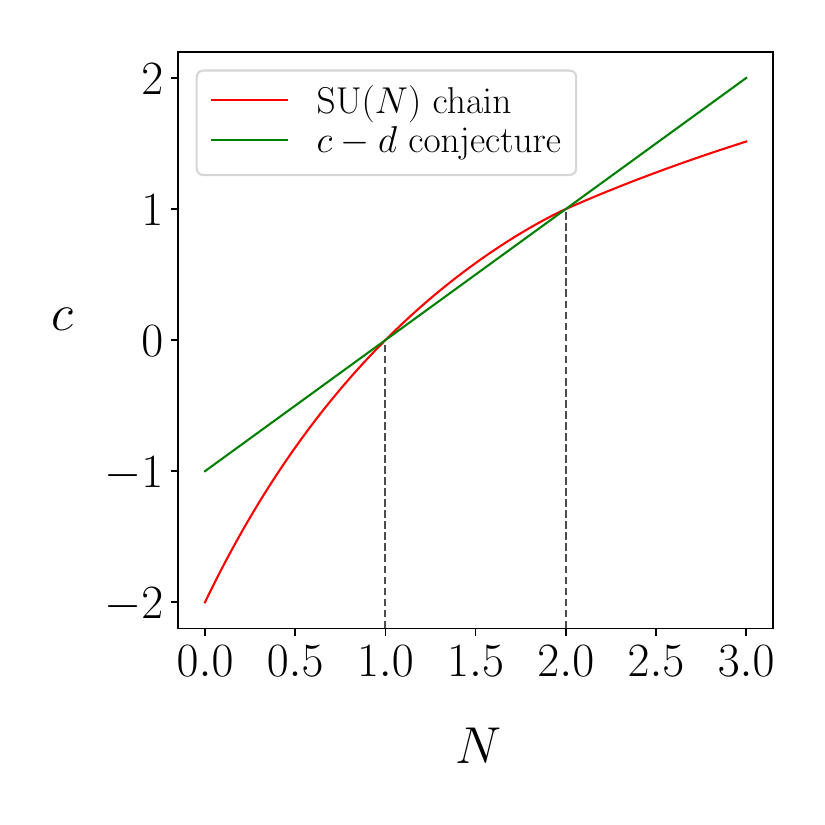}
    \caption{\label{fig:cd_bound} We plot the central charge $c$ of the SU($N$) Heisenberg chain (red line) along with the maximum allowed central charge predicted by the $c-d$ conjecture (green line). For $N \in (1, 2)$, the central charge exceeds the conjectured upper bound.}
\end{figure}

In a recent work~\cite{cd_conjecture}, Jos\'e I. Latorre and Germ\'an Sierra have conjectured that the maximum central charge ($c_{\text{max}}$) that a 1D critical local nearest neighbor Hamiltonian with the local Hilbert space dimension $d$ can have is $c_{\text{max}} \leq d-1$.
For the SU($N$) Heisenberg chain, the local Hilbert space dimension is $N$ and the central charge is bounded by $c_{\text{max}} \leq N-1$ as suggested by the $c-d$ conjecture.

In Fig.~\ref{fig:cd_bound}, we plot the central charge values for the SU($N$) chain given by $c = 1-\frac{6(g-1)^{2}}{g} \quad\text{with}\quad N=-2\cos(\pi g)$ ($g\in [\tfrac{1}{2},1]$) alongside the conjectured upper bound $c=N-1$. For $N>2$, where the central charge becomes complex, only its real part is shown.
We observe that for $N \in (1,2)$, SU(N) Heisenberg chain has central charge values bigger than the bound. We find this to be an interesting example that supports the $c-d$ conjecture in the unitary cases when $N=1,2$, and illustrates violations that can occur due to non-unitarity for non-integer $N$.

\subsection{Spin wave velocity (velocity of the excitations)}

Finite temperature QMC methods are based on mapping the partition function of a $d$-dimensional quantum system to a  $d+1$-dimensional classical system. At criticality, the imaginary time direction should be scaled (at least) in proportion to $v\beta$, where $\beta = \frac{1}{T}$ is the inverse temperature and $v$ is the velocity, called the spin wave velocity or the velocity of the excitations. Now, on this $d+1$-dimensional space, two kinds of topological numbers can be defined, temporal winding number ($W_{\tau}$) corresponding to the temporal direction and spatial winding numbers ($W_{r}$) corresponding to each spatial direction. The velocity of the excitations can be computed using the spatial and the temporal winding numbers~\cite{spin_wave_Jiang,spin_wave_Arnab}.

For a spin system, the square of the temporal winding number is related to the uniform magnetic susceptibility:

\begin{equation}
     \chi = \frac{\beta}{N}\langle M_{z}^{2} \rangle = \frac{\beta}{N} \langle W_{\tau}^{2} \rangle
\end{equation}

 and the square of the spatial winding number is related to the spin stiffness:
\begin{equation}
    \rho_{s} = \frac{1}{\beta}\Big( \frac{1}{d}\sum_{r=1}^{d}\langle W_{r}^{2}\rangle\Big)
\end{equation}

\begin{figure}[t]
    \includegraphics[width=1.0\linewidth, height = 8cm]{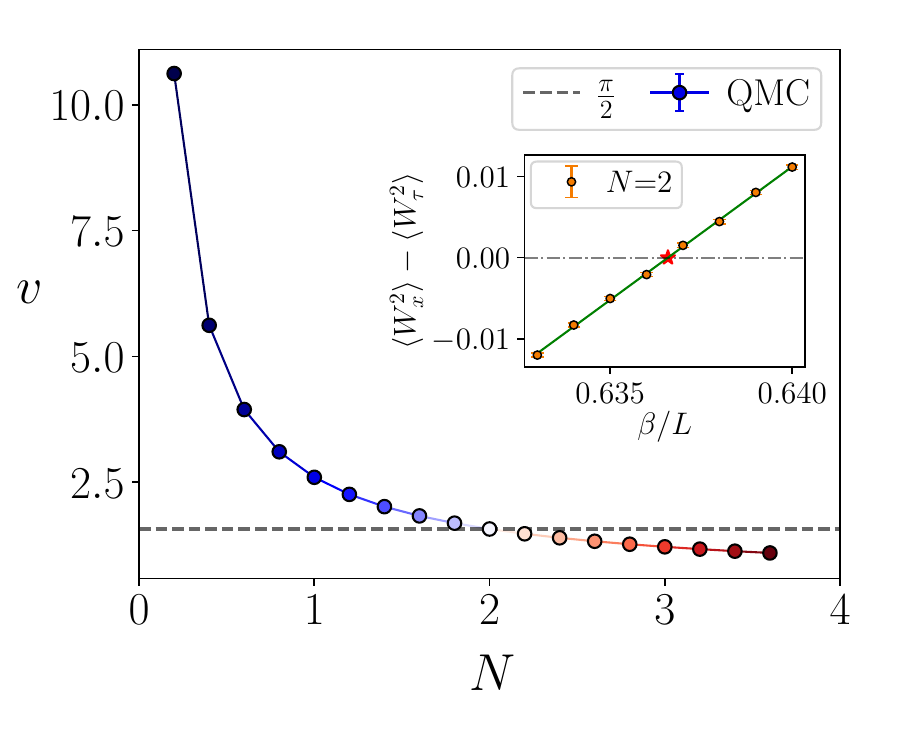}
    \caption{\label{fig:spin_wave}
    Spin wave velocity ($v$) as a function of $N$ for $N\in(0,3.6]$. The dashed horizontal line indicates the theoretical value of $v$ for $N=2$. For this value, we find $v=1.5708(1)$, which agrees with the theoretical prediction $\pi/2 = 1.5707$ within statistical errors. In the inset, we plot the difference between the square of the spatial winding number \( \langle W_x^2 \rangle \) and the square of the temporal winding number \( \langle W_\tau^2 \rangle \) as a function of \( \beta/L \) for \( N = 2 \) and \( L = 64 \). The orange data points are the QMC values, the solid green line is a linear fit, and the red star denotes the root $(\beta^{*}/L)$ obtained by solving the linear fit equation where \( \langle W_x^2 \rangle - \langle W_\tau^2 \rangle = 0 \). The inverse of this root, $v=L/\beta^{*}$,  gives the spin wave velocity.
    }
\end{figure}

For an unfrustrated antiferromagnetic system whose ground state is a singlet,  $\langle W_{\tau}^{2} \rangle \xrightarrow{} 0 $ in the limit $\beta \xrightarrow{} \infty $ ($T \xrightarrow{} 0 $). Similarly in the limit $\beta \xrightarrow{} 0 $ ($T \xrightarrow{} \infty $), we will have 
$\langle W_{\tau}^{2} \rangle > 0 $.
In the limit $\beta \xrightarrow{} \infty $ ($T \xrightarrow{} 0 $) a system with long-range order or quasiorder ( 1D system with power-law decaying correlations) has a nonzero stiffness constant for any $L$. So we will have $\langle W_{r}^{2} \rangle \sim \beta $ in this limit. In the limit, $\beta \xrightarrow{} 0 $ ($T \xrightarrow{} \infty $), the system will be in the disordered phase so we will have $\langle W_{r}^{2} \rangle \xrightarrow{} 0 $.
Due to the different limiting behaviors of the temporal and spatial winding numbers as a function of $\beta$, there will be a crossing point $\beta = \beta^{*}(L)$ for each $L$, where $\langle W_{\tau}^{2} (\beta^{*})\rangle  = \langle W_{r}^{2} (\beta^{*})\rangle$ and the finte-size estimate of the spin wave velocity is given by $v(L)=L/\beta^{*}$, which converges to the thermodynamic value in the limit $L\xrightarrow[]{} \infty$

In Fig.~\ref{fig:spin_wave}, we plot the computed spin wave velocities as a function of $N$ for the SU(N) Heisenberg chain.
We compute the $\langle W_x^2 \rangle$ and $\langle W_\tau^2 \rangle$ for a periodic chain of length $L=64$ (we find that $v(L)$ converges to the thermodynamic limit for the $L=64$) as a function of $\beta$ and then do a liner fit to determine the root ($\beta^{*}/L$) of the equation $\langle W_x^2 \rangle - \langle W_\tau^2 \rangle = 0 $ as shown in the inset for $N=2$. The inverse of this root, $v=L/\beta^{*}$,  gives the spin wave velocity. We find that $v$ decays exponentially as $N$ increases. For SU(2) critical antiferromagnetic Heisenberg chain, the value of $v$ is known theoretically to be equal to $\pi/2$~\cite{spin_wave_Arnab}. Our QMC value agrees with the theoretical prediction within statistical errors. For $N \neq 2$, theoretical values of $v$ are not known. For $N=3$, we have $v=1.216(1)$.

\begin{table*}[h!]
\centering
\setlength{\tabcolsep}{15pt} 
\renewcommand{\arraystretch}{2.0} 
\begin{tabular}{p{1.5cm}p{1.8cm}p{1.8cm}p{1.8cm}p{1.8cm}}
\toprule \toprule
$N$ & $x_{\text{CFT}}$ & $x_{\text{QMC}}$ & $c_{\text{CFT}}$ & $c_{\text{QMC}}$  \\ 
\midrule \midrule
$0.2$ & 1.820  & 1.85(7)  & -1.4719  & -1.475(1)     \\ \hline
$0.4$ & 1.659  & 1.67(6)  & -1.0210  & -1.0190(9)     \\ \hline
$0.6$ & 1.512  & 1.52(1)  & -0.6323  & -0.6297(2)    \\ \hline
$0.8$ & 1.377  & 1.39(1) & -0.2948  & -0.2934(1)    \\ \hline
$1.0$ & 1.25    & 1.25  & 0.0      &   0.0         \\ \hline
$1.2$ & 1.128  & 1.12(1)  & 0.2583   & 0.2568(2)     \\ \hline
$1.4$ & 1.008  & 1.00(1) & 0.4849   & 0.4812(5)    \\ \hline
$1.6$ & 0.886  & 0.89(1)  & 0.6834   & 0.6803(9)     \\ \hline
$1.8$ & 0.751  & 0.785(5)  & 0.8556   & 0.8497(5)     \\ \hline
$2.0$ & 0.5     & 0.59(1)  & 1.0       & 0.994(2)    \\ \hline
$2.2$ & 0.471+$i 0.208$  & 0.53(1)  & 1.117+$i 0.016$   & 1.117(3)     \\ \hline
$2.4$ & 0.443+$i 0.286$  & 0.414(3) & 1.226+$i 0.044$   & 1.213(3)    \\ \hline
\bottomrule 
\end{tabular}
\caption{Extracted central charge ($c_{\text{QMC}}$) and decay exponent ($x_{\text{QMC}}$) from fits to the Calabrese–Cardy scaling form [Eq.~\ref{eq.CC_form}] for $L = 48$ and $N \in [0.2, 2.4]$. For comparison, the corresponding CFT predictions, $x_{\text{CFT}}$ and $c_{\text{CFT}}$, are also shown. The extracted central charge values show excellent agreement with CFT predictions across all $N$. The decay exponent values also agree well, except near $N=2$, which are due to the known logarithmic corrections. The discrepancies for $N > 2$ may indicate a breakdown of the oscillatory form in Eq.~\ref{eq.CC_form} with complex scaling dimensions.
{\label{Table:N_2.4}}}
\end{table*}


\begin{table*}[h!]
\centering
\setlength{\tabcolsep}{12pt}
\renewcommand{\arraystretch}{1.1}
\begin{tabular}{p{2.5cm}p{2.5cm}p{2.5cm}}
\toprule \toprule
$N$ & $L$ & $c(L)$ \\
\midrule
\multirow{9}{*}{$2.6$} 
& 24  & 1.2999(8) \\
& 28  & 1.2992(7) \\
& 32  & 1.300(1) \\
& 40  & 1.290(2) \\
& 48  & 1.287(2) \\
& 56  & 1.276(3) \\
& 64  & 1.263(3) \\
& 72  & 1.258(4) \\
& 80  & 1.247(4) \\
\midrule
\multirow{9}{*}{$2.8$} 
& 24  & 1.3682(9) \\
& 28  & 1.3643(9) \\
& 32  & 1.355(1) \\
& 40  & 1.336(2) \\
& 48  & 1.315(3) \\
& 56  & 1.289(3) \\
& 64  & 1.265(4) \\
& 72  & 1.238(5) \\
& 80  & 1.208(7) \\
\midrule
\multirow{9}{*}{$3.0$} 
& 20  & 1.4580(4) \\
& 24  & 1.425(1) \\
& 28  & 1.411(3) \\
& 32  & 1.384(1) \\
& 40  & 1.337(2) \\
& 48  & 1.289(3) \\
& 56  & 1.235(4) \\
& 64  & 1.179(5) \\
& 72  & 1.130(8) \\
\midrule
\multirow{9}{*}{$3.2$} 
& 20  & 1.5090(8) \\
& 24  & 1.457(1) \\
& 28  & 1.430(1) \\
& 32  & 1.383(1) \\
& 40  & 1.304(2) \\
& 48  & 1.213(3) \\
& 56  & 1.117(4) \\
& 64  & 1.023(5) \\
& 72  & 0.909(8) \\
\midrule
\multirow{8}{*}{$3.4$} 
& 20  & 1.545(1) \\
& 24  & 1.472(1) \\
& 28  & 1.430(1) \\
& 32  & 1.359(2) \\
& 36  & 1.304(3) \\
& 40  & 1.22092) \\
& 44  & 1.161(4) \\
& 48  & 1.093(3) \\
\midrule
\multirow{7}{*}{$3.6$} 
& 20  & 1.567(1) \\
& 24  & 1.466(1) \\
& 28  & 1.40291) \\
& 32  & 1.297(2) \\
& 36  & 1.222(3) \\
& 40  & 1.114(3) \\
& 44  & 1.005(4) \\
\bottomrule\bottomrule
\end{tabular}
\caption{System size ($L$) dependent central charge values ($c_{L}$) for $N\in[2.6,3.6]$. We fit these to the central charge drift relation (Eq.~\ref{eq:c_drift}) to extract the real part of the complex central charge. 
{\label{Table:C_L}}}
\end{table*}

\begin{table*}[h!]
\centering
\setlength{\tabcolsep}{15pt} 
\renewcommand{\arraystretch}{2.0} 
\begin{tabular}{p{0.4cm}p{2.0cm}p{1.3cm}p{1.3cm}p{1.3cm}p{2.8cm}}
\toprule \toprule
$N$ & $c_{\text{CFT}}$ & $c_{R}$ & $\alpha$ &  $L_{0}$ & $\tilde{L_{0}}$ \\ 
\midrule \midrule
$2.6$ & 1.328+$i 0.079$   & 1.321(8)    & 0.006(4)  &  0.03(8) & 686603(1325758)
\\ \hline
$2.8$ & 1.424+$i 0.117$   & 1.43(1)     & 0.05(2)   &  2(1)    & 10120(3810)
\\ \hline
$3.0$ & 1.514+$i 0.157$   & 1.52(1)     & 0.21(3)   &  11(3)   & 2258(133)
\\ \hline
$3.2$ & 1.599+$i 0.199$   & 1.58(1)     & 0.33(1)   &  14(1)   & 1338(22)
\\ \hline
$3.4$ & 1.680+$i 0.243$   & 1.68(3)     & 0.39(4)   &  12(2)   & 908(30)
\\ \hline
$3.6$ & 1.756+$i 0.287$   & 1.74(6)     & 0.5(1)    &  13(4)   & 679(33) \\
\bottomrule \bottomrule
\end{tabular}
\caption{Extracted fit parameters from the central charge drift analysis for $N\in[2.6,3.6]$. For comparison, the corresponding CFT values for the complex central charges ($c_{\text{CFT}}$) are also shown. The fitted values of $c_{R}$ show excellent agreement with the real part of the complex central charge ($\Re(c_{\text{CFT}})$) across all $N$.
{\label{Table:N_3.6}}}
\end{table*}


\end{document}